\documentclass[twocolumn,tighten]{aastex63}
\usepackage{amsmath}

\received{}
\revised{}
\accepted{\today}
\submitjournal{AJ}

\shorttitle{NGC 6822 Kinematics}
\shortauthors{Belland et al.}

\graphicspath{{./}{figures/}}

\begin{document}

\title{NGC 6822 as a probe of dwarf galactic evolution\footnote{The
    data presented herein were obtained at the W.~M.~Keck Observatory,
    which is operated as a scientific partnership among the California
    Institute of Technology, the University of California and the
    National Aeronautics and Space Administration. The Observatory was
    made possible by the generous financial support of the W.~M.~Keck
    Foundation.}
}

\correspondingauthor{Brent Belland}
\email{bbelland@caltech.edu}

\author[0000-0003-1950-448X]{Brent Belland}
\affiliation{California Institute of Technology \\
1200 East California Boulevard \\
Pasadena, CA 91125, USA}

\author[0000-0001-6196-5162]{Evan Kirby}
\affiliation{California Institute of Technology \\
1200 East California Boulevard \\
Pasadena, CA 91125, USA}

\author[0000-0002-9604-343X]{Michael Boylan-Kolchin}
\affiliation{The University of Texas at Austin \\
2515 Speedway, Stop C1400 \\
Austin, TX 78712, USA}

\author{Coral Wheeler}
\affiliation{Carnegie Observatories  \\
813 Santa Barbara Street \\
Pasadena, California, 91101 USA}
\affiliation{NASA Hubble Fellow}

\begin{abstract}

NGC 6822 is the closest isolated dwarf irregular galaxy to the Milky Way.  Its proximity and stellar mass ($10^8 M_\odot$, large for a dwarf galaxy) allow for a detailed study of its kinematic properties. The red giant branch (RGB) stars at the galaxy's center are particularly interesting because they are aligned on an axis perpendicular to the galaxy's more extended \ion{H}{1} disk.  We detected a velocity gradient among the RGB population using spectra from Keck/DEIMOS\@.  This rotation is aligned with the \ion{H}{1} disk, but the sense of rotation is about the major axis of the central RGB population. We measured the rotation velocity ($v$) and velocity dispersion ($\sigma$) of the RGB population in five metallicity bins.  We found an increase of rotation support ($v/\sigma$) with increasing metallicity, driven primarily by decreasing dispersion. We also deduced an increasing radial distance for lower metallicity stars at $-0.5$~kpc/dex by relating the observed stellar kinematics to position via NGC 6822's \ion{H}{1} velocity curve. While the inverted metallicity gradient-like could be interpreted as evidence for an outside-in formation scenario, it may instead indicate that stellar feedback disturbed a centrally star forming galaxy over time.

\end{abstract}

\section{Introduction} \label{sec:introII}

There is a broad set of morphologies that describe massive galaxies \citep{2005ARA&A..43..581S}. Perhaps the best known morphology difference is between spiral and elliptical galaxies, as in Hubble's classification system \cite{1926ApJ....64..321H}. This visual distinction informs us about some of the physical properties of these galaxies: spiral galaxies contain neutral and ionized gas, actively form stars, and rotate, whereas elliptical galaxies are old, gas-poor, quiescent, and supported by random motion (i.e., dispersion).

Galaxies obey a morphology density relation \citep{1984ApJ...281...95P}. In denser environments, the population fraction of spiral galaxies decreases and the fraction of elliptical galaxies increases. This relation points to the formation mechanism of elliptical galaxies: In denser galactic environments, the spiral galaxies begin to merge. These mergers cause gas loss, star formation, and disruption of the rotation support.  The result is the transformation of a spiral galaxy into an elliptical galaxy.

For dwarf galaxies, there is a parallel between the spirals and ellipticals.  Dwarf irregulars are analogous to giant spirals because they are gas-rich, rotating, and star-forming.  Dwarf spheroidals are analogous to giant ellipticals because they are gas-poor, dispersion-supported, and quenched. Dwarf galaxies also obey their own morphology--density relation.  A dwarf galaxy's morphology correlated with the distance to its host galaxy. As found by \citet{2014ApJ...795L...5S}, dwarf galaxies within the Milky Way (MW) viral radius are predominantly gas-poor (i.e., dwarf spheroidal, dSph) while those just outside the MW viral radius are primarily gas-rich (i.e., dwarf irregulars, dIrr).

\cite{2001ApJ...559..754M} proposed a mechanism, called tidal stirring, to describe this morphology density relation. Successive pericentric passages near the MW can tidally disturb a dIrr. The ram pressure stripping and tidal influence of the MW cause angular momentum of the galaxy's gas to flow outward.  Then, the angular momentum is carried away when the outlying stars and gas are stripped from the galaxy. The loss of rotation and gas thus transitions the dIrr to a dSph and can explain the morphology--density relation.

However, in practice, the ``rotation support'' (ratio of rotation velocity divided by velocity dispersion) of the old populations in dIrrs is almost uniformly low and consistent with no rotation.  This quality holds for both observed galaxies and galaxies simulated by the Feedback in Realistic Environments \citep[FIRE,][]{2014MNRAS.445..581H} code \citep{2017MNRAS.465.2420W}. This is not a surprise for the dSphs, which are dispersion-dominated, but dIrrs are usually presumed to form with rotation.
\citet{2007MNRAS.382.1187K} note that, due to their lower mass, dwarf galaxies are more affected by pressure support from gas temperature, which causes them to be less disky and less rotation-supported than their more massive counterparts.  Nonetheless, some dIrrs clearly have rotating gas disks \citep[e.g.,][]{2017MNRAS.466.4159I}. What removes the rotation support from a dwarf irregular if not tidal stirring?

One of the most accessible dwarf irregular galaxies to study is NGC 6822. NGC 6822 is about 500 kpc from the MW, making it the nearest isolated dIrr. It is relatively massive, with $10^8 M_\odot$ in stars and a similar mass of \ion{H}{1} gas \citep{2012AJ....144....4M}. It's large angular extent (about $1.2^{\circ}$) permits studies at high angular resolution.  Spectroscopic studies enjoy the luxury of choosing individual stars and achieving high S/N for each star observed.

NGC 6822 has many interesting properties that contain clues about its history. 
NGC 6822 has an optically bright center, often referred to as a bar due to its elongated shape \citep{1996AJ....112.2596G}. As a result, NGC 6822 is classified as an irregular barred dwarf galaxy \citep{1976rcbg.book.....D}. Although the classification of the bar is based simply on the galaxy's visual appearance, \citet{2007ApJ...657..773V} raised the possibility that the central stellar population is a dynamical bar. However, there is also an \ion{H}{1} disk that runs normal to this bright optical center \citet{2000ApJ...537L..95D}. Ongoing star formation closely follows the shape of the disk \citep{2003ApJ...590L..17K}. There also is an older/intermediate-age stellar population associated with NGC 6822 that extends much farther than the bright optical center and does not follow the disk \citep{2002AJ....123..832L,2016MNRAS.462.3376T}.

On top of the stellar population distributions, the \ion{H}{1} gas disk has unique properties that are worthy of mention. There is a large \ion{H}{1} hole and in the southeast of the galaxy and a \ion{H}{1} overdensity in the northwest of the galaxy  \citep{2000ApJ...537L..95D}. \citet{2000ApJ...537L..95D} noted that the southwest of NGC 6822 had a tidal arm feature, indicating a previous interaction history. They estimated that the \ion{H}{1} hole had a relatively short kinematic age of 100 Myr, which could be correlated to the tidal arm feature if such an interaction increased the star formation rate. Noting the asymmetrically large mass distribution in the northwest of NGC 6822, they calculated a similar interaction timescale of 300 Myr if the overdensity in the northwest were a merger that had disrupted the southeastern hole, leading to a prediction that the northwest cloud was a companion galaxy.  \citet{2006AJ....131..343D} expanded on this interpretation, finding that the cloud was distinguishable from the rest of NGC 6822 in velocity and would have a dynamical mass ratio consistent with a dwarf galaxy merger.
However, \citet{2012ApJ...747..122C} later found that the putative companion had a similar star formation history as the rest of NGC 6822 and no old star overdensity or metallicity difference relative to the expected value from NGC 6822 at its distance. The stellar population was thus inconsistent with dwarf galaxy merger, even though they did not rule out a \ion{H}{1} cloud.

This paper addresses the formation history of NGC 6822 by analysis of the kinematics of the old, central population of stars.  Section~\ref{sec:Data} presents the already-published spectra and velocity measurements.  Section~\ref{sec:Results} gives our measurements of rotation and velocity dispersion in different bins of stellar metallicity.  We interpret these trends in Sec~\ref{sec:Discussion}, and we summarize in Section~\ref{sec:Summary}.

\section{Data} \label{sec:Data}

\begin{figure}[t]
\begin{center}
\includegraphics[width=\linewidth]{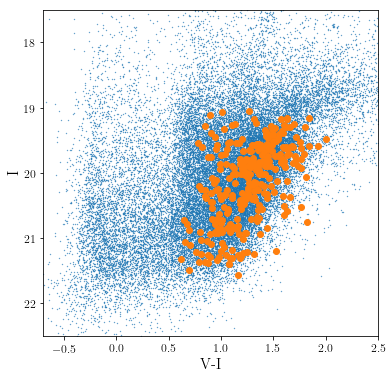}
\caption{Distribution of stars (orange) from \citet{2013ApJ...779..102K} in color--magnitude space plotted with photometry (blue) from \citet{2007AJ....133.2393M}.
\label{fig:CMD} }
\end{center}
\end{figure}

Kinematic data for NGC 6822 stars comes from the observations of \citet{2013ApJ...779..102K}. 299 stars between two slitmasks were observed in the Keck II telescope's Deep Imaging Multi-Object Spectrograph \citep[DEIMOS,][]{2003SPIE.4841.1657F} with total exposure times of 8.7 and 6.0 hours.  The color--magnitude diagram of stars in the \citeauthor{2013ApJ...779..102K} data is recreated in Figure~\ref{fig:CMD}.

Previous analysis of NGC 6822 by \citet{2014MNRAS.439.1015K} indicated that rotation might be present in NGC 6822 on the order of 10~km/s, but a definitive detection of rotation was uncertain due to the high velocity dispersion of $23.2\pm1.2$ km/s. \citet{2016MNRAS.462.3376T} later found rotation in carbon stars in a more extended population at $11.2\pm2.1$ km/s with a position angle of $26\pm13^\circ$. A similar rotation was also found in the gas (for example, see \citealt{2003MNRAS.340...12W}). 

The red giant branch star selection was done by selecting stars within a brightness, color, and isochrone cutoff. Target selection was completed using photometry data from \citet{2007AJ....133.2393M}. In this paper, the ages of stars are inferred from the metallicity of the stars, which itself is highly correlated with color. Therefore, any color cutoffs used in the star selection will impact the age/metallicity range considered. Stars with extinction-corrected magnitudes between 19.0 and 21.6 were selected based on the expected brightness of red giants at a distance modulus of 23.40 (479~kpc) \citep{2012MNRAS.421.2998F}. We applied a constant reddening correction of $E(B-V) = 0.25$ \citep{2007AJ....133.2393M}.  Stars beyond the colors between $0.6 < (V-I)_0 < 2.5$ were excluded based on the expected colors of red giants. Yonsei--Yale isochrones in $V$ and $I$ filters ranging from the bluest (2~Gyr, ${\rm [Fe/H]} = -3.76$) to the reddest (14~Gyr, metal-rich ${\rm [Fe/H]} = +0.05$) bounded the selection in the color--magnitude diagram. In cases where multiple stars competed for the same slit, priority was placed on stars closest to a 6~Gyr, ${\rm [Fe/H]} = -1.05$ isochrone.

Especially since this study relies on observing stars with a span of metallicity/ages, it is important to determine possible observation bias. The selection of red giant branch stars in \cite{2013ApJ...779..102K} was designed to be as inclusive as possible: the large isochrone metallicity range of -3.76 to +0.05 and large isochrone age range of 2-14 Gyr is expected to not reject any red giant branch star, thus not biasing the metallicity range of the stars in the sample. Due to the priority function increasing towards the intermediate age and metallicity isochrone, however, there may be a slight bias toward observing these intermediate age and metallicity stars. This bias is not expected to significantly affect the analysis in this paper, as while the priority function does complicate analysis of the metallicity bias selection, it only effects the uncommon cases where multiple stars were possible candidates for the same slit.

The stellar kinematics in a galaxy encode an important part of a galaxy's history. Stars are probes of the galactic structure as they travel through the galactic potential over time. Disruptive events such as mergers or substructure such as tidal streams can be tracked by following stellar groups that are distinct in metallicity and velocity space.  Rotation is commonly a criterion referenced to distinguish between dwarf spheroidal and dwarf irregular galaxies with dwarf spheroidals expected to have undergone a significant morphological transition \citep{2001ApJ...559..754M}.

We examine the kinematics of NGC~6822 through existing measurements of the radial velocities of red giants \citep{2014MNRAS.439.1015K}.  Because \citeauthor{2014MNRAS.439.1015K}\ restricted their measurements to red giants, our study is sensitive to stellar populations older than $\sim 1$~Gyr.  Younger stars, like blue and red supergiants and stars on the upper main sequence, are not present in our sample.  Furthermore, the sample is restricted to the central region of the galaxy.  Specifically, the measurements are confined to an approximately rectangular region with an area of about $8' \times 15'$.  The red giant population extends significantly beyond this region \citep{2012ApJ...747..122C}.

\section{Results} \label{sec:Results}

\subsection{Prolate rotation}

\begin{figure}[t]
\begin{center}
\includegraphics[width=\columnwidth,angle=0]{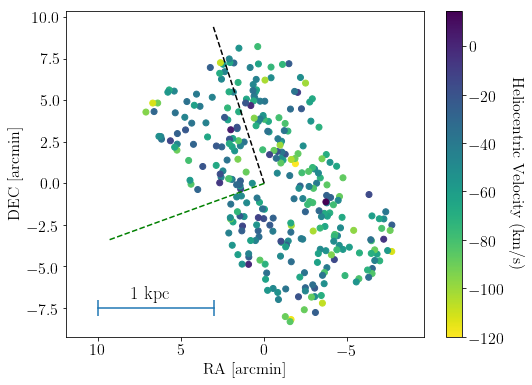}
\caption{Distribution of stars from \citet{2013ApJ...779..102K}, color-coded by heliocentric velocity, with dark blue corresponding to least rapidly approaching and light yellow corresponding to most rapidly approaching.   The dotted black line indicates an approximate position angle of the major axis of RGB stars estimated from the RGB isophotes in Figure~8 of \citet{2012ApJ...747..122C}  (18 degrees), while the green line indicates the approximate position angle of the major axis of \ion{H}{1} disk from \citet{2003MNRAS.340...12W} (110 degrees). The velocity gradient apparent from west to east indicates rotation, prolate in the sense of the RGB stars and oblate in the sense of the \ion{H}{1} disk.  Coordinates are in units of arcmin relative to the center of NGC 6822 \citep{2012AJ....144....4M}. The angular size of 1 kpc is shown in the bottom left for reference; NGC 6822 extends 5 kpc in either direction along the \ion{H}{1} major axis and about 1 kpc in either direction along the RGB isophote major axis. 
\label{fig:Rot} }
\end{center}
\end{figure}

Figure~\ref{fig:Rot} presents the map of radial velocities of red giants in the center of NGC~6822. Despite the large scatter in stellar
velocities, there is a clear velocity gradient across NGC 6822, which shows that the red giant population is rotating. The magnitude of this rotation is about $\pm 10$ km/s, which is about half of the velocity dispersion of
these stars \citep{2014MNRAS.439.1015K}. The ratio between the rotation velocity and velocity dispersion indicates the balance of different mechanisms that support the galaxy against gravity. This ratio is about
0.5 for these stars but notably greater than zero.

This nontrivial rotation of the old red giant branch (RGB) stars in Figure~\ref{fig:Rot} is
around the major axis of the innermost RGB stars, as estimated from Figure 8 of \citet{2012ApJ...747..122C}. (\citet{2006A&A...451...99B} observed a position angle of about $65^\circ$ for RGB stars that extend for the entirety of NGC 6822.  This angle does not represent the RGB stars in the central core of the galaxy.) Rotation about a major axis is an uncommon phenomenon; most rotation
is about a minor axis, such as a spiral galaxy rotating around its minor axis
rather than twirling like a flipped coin through space. In fact, prolate rotation is so unusual that it has even been invoked as
possible evidence of a merger history for the Andromeda II \citep{2012ApJ...758..124H} and Phoenix \citep{2017MNRAS.466.2006K} dwarf galaxies.

If NGC 6822's RGB stars are prolately rotating, this
may thus be evidence that NGC 6822 has experienced a merger in its past.
This would not be the first time a merger history has been considered for this
galaxy; \citeauthor{2000ApJ...537L..95D}\ independently proposed a merger history for NGC 6822 due to its morphology and jump in \ion{H}{1} disk velocity.

The apparent prolate rotation of the RGB stars is informative even if there was not a merger in NGC 6822's past. If NGC 6822's old population is prolately
rotating due to internal processes rather than external mergers, then a dwarf
irregular like NGC 6822 could transition into a prolately rotating dwarf
spheroidal from just removal of its gas. \citet{2017MNRAS.465.2420W} similarly argued
that the low rotation support of the RGB star population in dIrrs could mean that transitioning to dSphs could be done simply by gas removal.

However, the rotation of the RGB stars as seen in Figure~\ref{fig:Rot} is not just
around their own major axis, but also along the major axis of the \ion{H}{1} disk
\citep{2003MNRAS.340...12W}. Rotation about the same axis as the \ion{H}{1} disk makes
sense if the RGB stars initially formed from the gas in the disk. However,
as the distribution of the central RGB stars doesn't resemble that of the disk, it is
apparent that there is a driver of morphological change between the disk and
the central RGB stars. 

Understanding this transition will help in understanding how dwarf galaxies like NGC 6822 evolve over time.

\subsection{Kinematics evolution}

There is clearly a change in the kinematic character between the \ion{H}{1} disk and the RGB stars, so the galaxy must have evolved over time. However, how does this transition occur? Was it sudden, such as from a quick merger, or an internal process over a long period of time? 

One common diagnostic of the kinematic evolution of a galaxy is an age--velocity dispersion relation (AVR) (for an example, see \citealt{2017MNRAS.472.1879L}). In order to approach this problem similarly, we also need to measure the ages and velocity dispersion of our sample of stars.

By classifying stars by age and determining how their kinematics vary as a function thereof, we can probe the transition between the effectively zero-age population of the galaxy (\ion{H}{1} gas) and the oldest stellar population of RGBs. One effective method for characterizing ages of stars is to calculate their metallicities: because galaxies increase in metallicity over time, stars with lower metallicity also tend to have been formed earlier. The correlation between age and metallicity can be complicated by stellar migration, as stars forming in separate regions with differing metallicity growth can be mixed by migration, creating metallicity differences in local populations independent of age. While outside the scope of the paper, determination of alpha-to-iron abundances in stars can disambiguate the formation environment of a star from age evolution, which would reduce uncertainty in this correlation. In the case of NGC 6822, \cite{2001AJ....122.2490W} found that metallicity is very nearly monotonic with age (Figures 19--22), adding reassurance that metallicity as a proxy for age is a valid assumption.

In addition to age, velocity dispersion information must be considered. Perturbations to a star's orbit are encoded in its velocity dispersion, which gives information about the rotational support of the gas these stars formed from.  Alternatively, the star's orbit can be shaped by disruptive events (kinematic collisions or mergers) or a disruptive potential of the host galaxy. In order to conserve total energy, dispersion of a population also increases if rotation velocity decreases. Given this relation between dispersion and rotation, the ratio of rotational velocity divided by velocity dispersion can be used to understand how the rotational support changes between the gas and stars.

In order to calculate rotational velocity and velocity dispersion as a function of time, each of the $j$ stars with metallicities within a given range (quintile of [Fe/H]) were grouped together in a bin. Within each of the $k$ bins, velocity dispersion $\sigma$ within each bin was determined from the line-of-sight velocities $v$ and uncertainties $\epsilon$ from finding the maximum of the likelihood function $L_k$ (as in \citealt{2017MNRAS.465.2420W}):

\begin{align}
    L_k = \prod_{j=1}^{N}\Bigl(&\frac{1}{\sqrt{2\pi(\sigma_k^2+\epsilon_j^2)}} \nonumber \\ & \times \text{exp}{\left[-\frac{1}{2}\frac{(v_j-(\bar{v}+v_{{\rm rot},k}\cos{(\theta_k-\theta_j)}))^2}{\sigma_k^2+\epsilon_j^2}\right]}\Bigr)\label{eq:like}
\end{align}
where $\bar{v}$ is the mean velocity of all stars in the galaxy, $\theta_k$ is the angle of the rotation axis for a bin, $v_j$ is the line-of-sight velocity of star $j$ belonging to bin $k$, and $\theta_j$ is the angle from the center of the galaxy to that star.

From the exponential of equation~\ref{eq:like}, the velocity rotation model adopted for a star is the average velocity plus a constant rotation velocity ($v_{{\rm rot},k}$) multiplied by an angular projection from the direction of rotation. \citet{2017MNRAS.465.2420W} found that the constant rotation model with just parameter $v_{\rm rot}$ was preferred to their pseudo-isothermal sphere model and is thus used here.  

The product in Equation~\ref{eq:like} combines all of the fits for each star to maximize simultaneously. The free parameters are the  $v_{{\rm rot},k}$, $\sigma_k$, $\bar{v}$ and $\theta_k$.  They were found through maximum likelihood using an MCMC with a Metropolis algorithm and $10^5$ links in the chain.  A burn-in period of $10^3$ links was discarded from the beginning of the chain.  This burn-in was found by visual inspection to be sufficient to decouple the chain from the initial choices of free parameters.

\begin{figure}[t]
\begin{center}
\includegraphics[width=\columnwidth,angle=0]{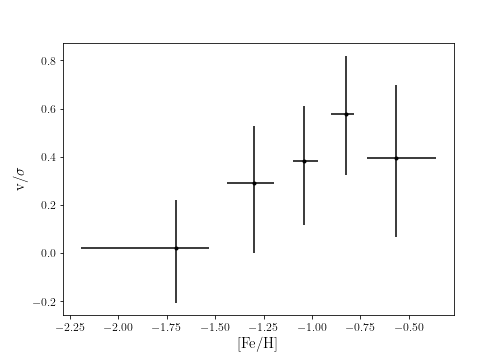}
\caption{Metallicity versus rotation support (rotational velocity divided by velocity dispersion) for stars in NGC 6822 separated into five equal-number bins. The errors bars represent 68\% confidence intervals.  The upward trend in rotation support with metallicity indicates that younger stars are more rotationally supported in NGC 6822. This information is included in the summary of Table~\ref{tab:1}.
\label{fig:MVR} }
\end{center}
\end{figure}

With the velocity and metallicity data, plotting the metallicity versus $v_{{\rm rot},k}/\sigma_k$ (``rotation support'') in each bin results in Figure~\ref{fig:MVR}. Notably, there in an increase in the rotation support in NGC 6822 with increasing metallicity, which corresponds to an increased rotation support for younger stars.  This behavior is expected if the RGB stars were disrupted from a rotationally dominated disk, or if the rotation support grew over time.

\section{Discussion} \label{sec:Discussion}
The evolution of kinematics with metallicity is a tracer of how the galaxy evolved over time, assuming that metallicity increases monotonically with time. However, separating out separate age populations also has a benefit of probing the gravitational potential of a galaxy with separate populations that have their own kinematic properties. \citet{2011ApJ...742...20W} effectively utilized this property by extracting out enclosed mass from two independent chemodynamic populations at two separate radii from the Fornax and Sculptor dwarf spheroidal galaxies.  Although the two populations trace the same underlying potential, they have separate velocity dispersions and separate physical sizes.  Therefore, these two measurements of enclosed mass reveal the slope of the enclosed mass profile. This mass slope can be used to distinguish between cuspy and cored mass profiles, which itself distinguishes between models of galaxy formation \citep[e.g.,][]{1996ApJ...462..563N,1997ApJ...490..493N,2008Sci...319..174M}. Thus, it is clear that the kinematics within multiple-age populations is a powerful probe of galactic evolution.

\begin{figure}[t]
\begin{center}
\includegraphics[width=\columnwidth,angle=0]{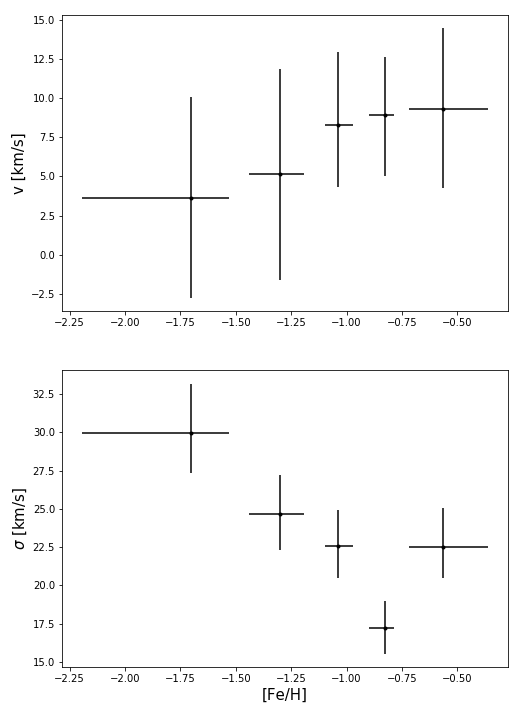}
\caption{Top: Rotation velocity vs.\ metallicity. Bottom: Velocity dispersion vs.\ metallicity. Taken together, the rotation velocity and velocity dispersion are the components of Figure~\ref{fig:MVR}. The velocity dispersion change is more significant than the rotation velocity. The errors bars represent 68\% confidence intervals.  The upward trend in rotation support with metallicity indicates that more metal-rich, presumably younger stars are more rotationally supported in NGC 6822. This information is included in the summary of Table~\ref{tab:1}.
\label{fig:RotDisp} }
\end{center}
\end{figure}

NGC 6822 has increasing rotation support with increasing metallicity (Figure~\ref{fig:MVR}). However, the large uncertainties due in part to the weak rotation and the large velocity dispersion obfuscate any detailed structure in the stellar rotation curve.  To understand Figure~\ref{fig:MVR} and the kinematics of NGC 6822 in more detail, we split the rotation support into contributions from rotation and from dispersion in Figure~\ref{fig:RotDisp}.  Dispersion notably decreases with increasing metallicity, as noted by \citet{2016MNRAS.456.4315S}. That study separated the galaxy into two populations: metal-rich and metal-poor.  Our increased sample size allows us to split the galaxy into five metallicity bins.  Our finer view of the evolution of dispersion shows that dispersion steadily decreases across a range of metallicity except for the most metal-rich bin.

The kinetic energy in any metallicity bin is a combination of the energy in rotation and energy in dispersion. We estimated the line-of-sight rotation velocity $v_k$ using the likelihood function in Equation~\ref{eq:like}, but the rotation is in the plane of the galaxy, which is inclined relative to the observer. The in-plane rotation velocity is $\frac{v_k}{\sin{i}}$, and rotational energy per unit mass is $\frac{1}{2}\left(\frac{v_k}{\sin{i}}\right)^2$. The inclination $i$ of NGC 6822 is 60 degrees \citep{2003MNRAS.340...12W}. The line-of-sight dispersion is $\sigma_k$ also found from the likelihood in Equation~\ref{eq:like}.

Assuming dispersion is isotropic, kinetic energy per unit mass from the 3D velocity dispersion would be $3\cdot\frac{1}{2}\sigma_k^2$ in a given bin. However, stellar kinematics are more affected by anisotropy for our centrally concentrated sample of stars compared to the whole stellar population. That is, if dispersion were only radial, line-of-sight dispersion at the center of the galaxy would encapsulate all energy in dispersion, which would be $\frac{1}{2}\sigma_k^2$. However, a primarily tangential dispersion would indicate that observations of line-of-sight velocities toward the center of the galaxy vastly underestimate the kinetic energy in the stars. To accommodate this uncertainty, we write kinetic energy in dispersion as $n\cdot\frac{1}{2}\sigma_k^2$, where $n=1$ corresponds to a purely radial dispersion, $n=3$ corresponds to isotropic dispersion, and $n>3$ corresponds to more tangentially dominant dispersion.

\begin{figure}[t]
\begin{center}
\includegraphics[width=\columnwidth,angle=0]{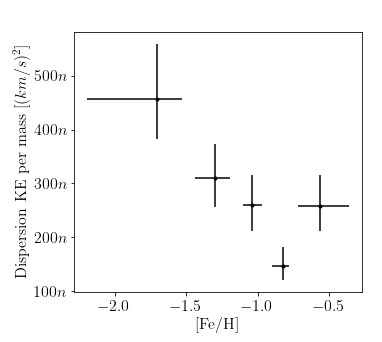}
\caption{Kinetic energy in dispersion per unit mass of stars in each metallicity bin. The kinetic energy trends in the same way as the dispersion (see Figure~\ref{fig:RotDisp})  because dispersion dominates over rotation. The variable $n$ quantifies uncertainty in anisotropy (see text). Notably, kinetic energy is not constant with metallicity. This information is included in the summary of Table~\ref{tab:1}.
\label{fig:kinetic} }
\end{center}
\end{figure}

For stars distributed over some radius $R_{\textbf{max}}$, this kinetic energy is associated with the gravitational potential energy at the average radius, on the order of $\frac{1}{2} R_{\textbf{max}}$. However, the exact value of radius depends on the system configuration and anisotropy. More specifically, \citet{2010MNRAS.406.1220W} found that enclosed mass at the half-light radius scales as $3 G^{-1} \sigma_k^2 r_{1/2}$.  There is an enclosed mass dependence on velocity anisotropy, $\beta$ which is minimized at the 3D (deprojected) half-light radius.  In total, kinetic energy at a given metallicity is given by $\frac{1}{2}\left(\frac{v_k}{\sin{i}}\right)^2 + \frac{n}{2}\sigma_k^2$. The kinetic energy is plotted in Figure~\ref{fig:kinetic}.

Unsurprisingly, the dispersion component dominates the weak rotation component, so the kinetic energy resembles the dispersion component of Figure~\ref{fig:RotDisp}. Since energy must be conserved but kinetic energy decreases with increasing metallicity, the effective potential energy per unit mass $-\frac{GM_{\text{enclosed}}(r)}{r}$ must be increasing with metallicity to balance this effect out. The radius $r$ is measured from the galactic rotation axis, and $M_{\text{enclosed}}(r)$ is the total mass enclosed within this radius. 
The total energy per unit mass is

\begin{equation}
    \frac{1}{2}\left(\frac{v_k}{\sin{i}}\right)^2 + \frac{n}{2}\sigma_k^2 -\frac{GM_{\text{enclosed}}(r)}{r} < 0 \label{eq:energy}
\end{equation}

\noindent and by the virial theorem, twice the kinetic energy of the stars on average should equal the negative of the potential energy:

\begin{equation}
    \left(\frac{v_k}{\sin{i}}\right)^2 + n\sigma_k^2 -\frac{GM_{\text{enclosed}}(r)}{r} = 0 \label{eq:virial}
\end{equation}

Changes in kinetic energy thus allow measurement of the change in average $-\frac{M_{\text{enclosed}(r)}}{r}$ between metallicity bins. Notably, the increase in $-\frac{GM_{\text{enclosed}(r)}}{r}$ with metallicity indicates that the average $r$ of each population must decrease with increasing metallicity. That is, metal-rich populations are more centrally concentrated. Such a relation was predicted by \citet{2016MNRAS.456.4315S} and can be seen in the young versus old stellar population distributions of \citet{2006AJ....131..343D}.

\begin{figure}[t]
\begin{center}
\includegraphics[width=\columnwidth,angle=0]{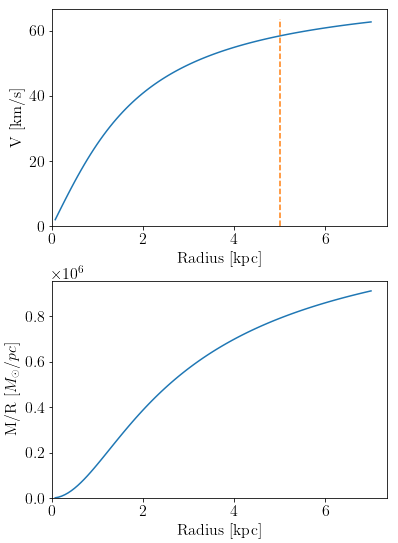}
\caption{Top: Approximate velocity versus radius from the B24 minimum disk model of \citet{2003MNRAS.340...12W}. The dotted orange line indicates the radial limit of the \citet{2003MNRAS.340...12W} data, with an extension from the pseudo-isothermal model in velocity beyond this value. Bottom: Conversion to enclosed mass divided by radius as a function of radius (i.e., Equation~\ref{eq:virial} where $\sigma_k = 0$).
\label{fig:HItoM} }
\end{center}
\end{figure}

\begin{figure}[t]
\begin{center}
\includegraphics[width=\columnwidth,angle=0]{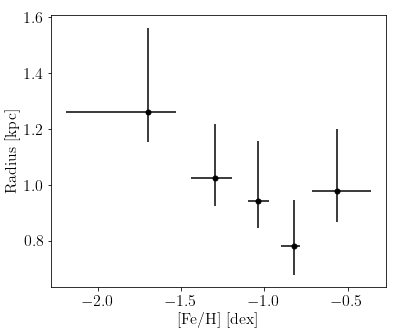}
\caption{Radius vs.\ Metallicity, combining the enclosed mass vs.\ metallicity relation (Figure~\ref{fig:kinetic}) with the velocity data of \citet[][Fig.~\ref{fig:HItoM}]{2003MNRAS.340...12W}, assuming the stellar dispersion is purely radial. Note that this is not the same as the metallicity gradient (see text). This information is included in the summary of Table~\ref{tab:1}.
\label{fig:FeHR_radial} }
\end{center}
\end{figure}

\begin{figure}[t]
\begin{center}
\includegraphics[width=\columnwidth,angle=0]{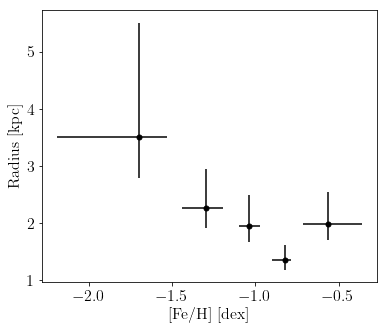}
\caption{Same as figure~\ref{fig:FeHR_radial} except assuming the stellar dispersion is isotropic. The decreasing velocity curve in HI indicates much further radial distances for metal-poor populations in this case. This information is included in the summary of Table~\ref{tab:1}.
\label{fig:FeHR_isotropic} }
\end{center}
\end{figure}

Alternatively, the enclosed mass divided by radius can be independently determined from the \ion{H}{1} gas velocity curve as a function of radius.  A pseudo-isothermal velocity curve model used by \citet{2003MNRAS.340...12W} to model the \ion{H}{1} velocity curve is shown in Figure~\ref{fig:HItoM}. Combined with the enclosed mass vs.\ metallicity relationship (Figure~\ref{fig:kinetic}), the spatial distribution of the different stellar populations can be probed. The inferred distributions for a purely stellar radial dispersion ($n=1$) and isotropic dispersion ($n=3$) are shown in Figures~\ref{fig:FeHR_radial} and~\ref{fig:FeHR_isotropic}.

\begin{deluxetable*}{lllllll}
\tablecaption{Table summarizing the stellar binned data in Figures \ref{fig:MVR}, \ref{fig:RotDisp}, \ref{fig:kinetic}, \ref{fig:FeHR_radial}, and \ref{fig:FeHR_isotropic}.\label{tab:1}}
\tablehead{
\colhead{[Fe/H] (dex)} & \colhead{Number of Stars} & \colhead{$v$ (km/s)} & \colhead{$\sigma$ (km/s)} & \colhead{Dispersion K.E. per mass $(km/s)^2$} &\colhead{Radius, $n=1$ (kpc)}  & \colhead{Radius, $n=3$ (kpc)}
}
\startdata
$-1.70^{+0.17}_{-0.49}$ & 60 & $3.6 \pm 6.4$ & $30.0^{+3.2}_{-2.6}$ & $449n^{+100n}_{-75n}$ & $1.3^{+0.3}_{-0.1}$ & $3.4^{+1.9}_{-0.7}$\\
$-1.30^{+0.11}_{-0.14}$ & 60 & $5.2 \pm 6.7$ & $24.6^{+2.5}_{-2.3}$ & $303n^{+67n}_{-54n}$ & $1.0^{+0.2}_{-0.1}$ & $2.2^{+0.7}_{-0.3}$\\
$-1.04^{+0.07}_{-0.06}$ & 58 & $8.3^{+4.7}_{-3.9}$ & $22.5^{+2.4}_{-2.1}$ & $254n^{+57n}_{-44n}$ & $0.95^{+0.18}_{-0.10}$ & $1.9^{+0.5}_{-0.3}$\\
$-0.82^{+0.04}_{-0.08}$ & 60 & $8.9^{+3.9}_{-3.7}$ & $17.2^{+1.8}_{-1.6}$ & $148n^{+33n}_{-27n}$ & $0.75^{+0.15}_{-0.09}$ & $1.3^{+0.3}_{-0.2}$\\
$-0.56^{+0.20}_{-0.15}$ & 61 & $9.3^{+5.0}_{-5.2}$ & $22.5^{+2.5}_{-2.1}$ & $254n^{+60n}_{-45n}$ & $0.97^{+0.21}_{-0.11}$ & $2.0^{+0.6}_{-0.3}$\\
\enddata
\end{deluxetable*}

As qualitatively predicted earlier, more metal-rich star populations are found toward the center of NGC 6822, and more metal-poor stellar populations have larger effective radii. These observations are consistent with the outside-in star formation processes found in other dwarf galaxies, where star formation became more centrally concentrated over time.
Alternatively, these negative metallicity gradients have been found to form in dwarf galaxies ($M_* \sim 10^{7-9.6}~M_{\sun}$) by \citet{2016ApJ...820..131E,2018MNRAS.473.1930E} due to feedback from star formation that drives fluctations in the galactic potential.  These perturbations cause kpc-scale migrations that cause older stellar populations to migrate outward and generate a negative metallicity gradient. Given that NGC 6822 neatly falls within this mass range, and due to its on-off star formation episodes \citep{2001AJ....122.2490W,2014ApJ...789..147W} that correlate to this model, such an inside-out model may be preferred to explain the more radially extended low-metallicity population in NGC 6822.

The radial distribution of stars with metallicity is about $-2$ kpc/dex in the isotropic case and $-0.5$ kpc/dex in the radial dispersion case. The lower value is not surprising in the radial case because radial outflows would more efficiently disperse older stars over time. Furthermore, the distances required in the radial dispersion case are more realistic, being more closely confined the the center of the galaxy than in the isotropic case due to less total energy in the stellar populations. In fact, active star formation histories in the centers of dwarf galaxies have been shown to induce radial anisotropy \citep{2016ApJ...820..131E}.  These calculated radii indicate a large radial isotropy consistent with an active star formation history. Such a process will be considered more in the next section.

Note that we averaged populations into metallicity bins, and we found the average radii of these bins. In other words, these measurements with units of kpc/dex indicate how populations at lower metallicities tend to be further away from the center of the galaxy. This quantity differs from a metallicity gradient (dex/kpc), which bins together stars at the same radius and then calculates average metallicity at each bin. That is, while the reciprocals of the calculated values indicate there is a negative metallicity gradient, which is considered in the rest of the analysis of this paper, they cannot be directly compared to metallicity gradients in NGC 6822 \citep[e.g.,][who found that metallicity dispersion prevented a precise analysis]{2016MNRAS.456.4315S}. 

\subsection{Formation History} 
 
We can also consider why the metal-rich, young stars are more centrally concentrated. Unlike in spiral galaxies where star formation occurs in the arms rather than the center, the younger star distribution in NGC 6822 is centrally concentrated. However, many dSphs have been found to have centrally concentrated younger stellar populations. 

One possibility for a negative metallicity gradient is that stars in NGC 6822 formed in a disk, but star formation in outer regions of the galaxy decreased more rapidly with time, due to reasons such as ionization of the outer \ion{H}{1} disk \citep{2006ApJ...641..785K} or merely gas depletion. This can explain why dSphs stop forming stars in their outskirts first, yet the dwarf irregular NGC 6822 still has plenty of gas. This does not explain why the the youngest, most-metal rich stars are at a slightly larger radius than the second most metal-rich bin.

Another possibility is that a merger event occurred, funneling new gas toward the center of the galaxy while also relocating old stellar populations to larger radii. NGC 6822 has been speculated to have a merger event in its history, invoked by \citet{2000ApJ...537L..95D} to explain NGC 6822’s \ion{H}{1} velocity curve. Such an event may be expected to bifurcate the stellar population, with separable old and young populations, whereas the observed metallicity gradient in NGC 6822 seems to be more continuous between bins. However, it may be possible for smaller mergers over time to smooth out such a gradient \citep{2016MNRAS.456.1185B}.

One possible explanation for the distribution of the RGB stars appearing to rotate in a prolate sense is that these stars trace a dynamical bar along the line of sight. If such a bar existed, it would affect the dynamics of the rest of the galaxy, perhaps in a noticeable way. In a set of simulations by \citet{1994ApJ...430L.105F}, abundance gradients in barred galaxies were found to be flattened with a possibly metal-rich center. This prediction does not match with the the observed metallicity gradient for NGC 6822 (e.g. Figure~\ref{fig:FeHR_isotropic}), indicating that the RGB population may not occupy a bar. However other studies, such as \citet{2009A&A...495..775P,2019MNRAS.483.1862Z}, found a more complicated family of possibilities, with positive, zero, and negative metallicity gradients in barred galaxies. The negative metallicity gradients (corresponding to NGC 6822's case) were explained as originating from the disk not having enough time to flatten out. Such negative metallicity gradients also tend to correlate with positive age gradients (younger populations toward the center).  This metallicity gradient relation is intriguing, but it does not appear to distinguish between other possible formation scenarios.

Radial migration of stars has also been used to explain metallicity gradients in galaxies: unordered radial migration may dilute an initially strong metallicity gradient. Radial migration can be induced by mergers, perturbations by companion galaxies, or galactic substructures. \citet{2016ApJ...818L...6L} were able to explain the varied metallicity distributions in the Milky Way as a function of radius by following radial migration in simulations. High-metallicity stars formed toward the center of the galaxy but migrated outward, enriching and skewing the outer radial metallicity distributions. In dwarf galaxies, an independent and dominant driver of radial migration may be induced by feedback during bursty star formation histories. \citet{2016ApJ...820..131E} found that in $M_* = 10^{7-9.6}~M{\sun}$ dwarf galaxies, feedback disrupting star formation in the center of the galaxy systematically drove out stars toward the galactic edges, creating metallcity gradients similar to that observed in our data for NGC 6822. Notably, such a high central star formation history would induce large, preferentially radial inflows and outflows of stars and gas during star formation bursts, which would also be consistent with the kinematic results of the previous section.

\section{Summary} \label{sec:Summary}

In conclusion, NGC 6822 is a promising candidate to study dwarf irregular populations and dynamics due to its mass and proximity. RGB archival data from \citet{2013ApJ...779..102K} revealed a 10 km/s rotation in the center of the galaxy, approximately half that of the dispersion. This rotation is unusual in that it is prolate relative to the distribution of RGB stars, though it also matches the rotation of the \ion{H}{1} disk. Due to the large number of stars and precise metallicity of the data, the rotation and velocity dispersion of the RGB population were measured in five metallicity bins. The kinematics across the bins were compared to the gas, which probes the same potential as the stars, to determine the radial extent of each population. A radial gradient of $-0.5$~kpc/dex (in the radial case) was found by relating the observed stellar kinematics to position via NGC 6822's \ion{H}{1} velocity curve. Negative metallicity gradients are sometimes correlated with outside-in star formation in a galaxy. However, NGC 6822's multiple episodes of star formation may instead indicate that stellar feedback induced migration of preferentially older stars out of the center of the galaxy.

\acknowledgements
The authors thank the anonymous referee for their input which improved this paper.

This material is based upon work supported by the National Science
Foundation under grant Nos.\ AST-1636426 and AST-1847909.  E.N.K.\ gratefully
acknowledges support from a Cottrell Scholar award administered by the
Research Corporation for Science Advancement as well as funding from
generous donors to the California Institute of Technology. MBK acknowledges support from NSF CAREER award AST-1752913, NSF grant AST-1910346, NASA grant NNX17AG29G, and HST-AR-15006, HST-AR-15809, HST-GO-15658, HST-GO-15901, and HST-GO-15902 from the Space Telescope Science Institute, which is operated by AURA, Inc., under NASA contract NAS5-26555. Support for CW was provided by NASA through the NASA Hubble Fellowship grant HST-HF2-51449.001-A, awarded by the Space Telescope Science Institute, which is operated by the Association of Universities for Research in Astronomy, Inc., for NASA, under contract NAS5-26555.

We are grateful to the many people who have worked to make the Keck
Telescope and its instruments a reality and to operate and maintain
the Keck Observatory.  The authors wish to extend special thanks to
those of Hawaiian ancestry on whose sacred mountain we are privileged
to be guests.  Without their generous hospitality, none of the
observations presented herein would have been possible.

\facility{Keck:II (DEIMOS)}

\bibliography{NGC6822}{}

\begin{thebibliography}{}
\expandafter\ifx\csname natexlab\endcsname\relax\def\natexlab#1{#1}\fi
\providecommand{\url}[1]{\href{#1}{#1}}
\providecommand{\dodoi}[1]{doi:~\href{http://doi.org/#1}{\nolinkurl{#1}}}
\providecommand{\doeprint}[1]{\href{http://ascl.net/#1}{\nolinkurl{http://ascl.net/#1}}}
\providecommand{\doarXiv}[1]{\href{https://arxiv.org/abs/#1}{\nolinkurl{https://arxiv.org/abs/#1}}}

\bibitem[{{Battinelli} {et~al.}(2006){Battinelli}, {Demers}, \&
  {Kunkel}}]{2006A&A...451...99B}
{Battinelli}, P., {Demers}, S., \& {Kunkel}, W.~E. 2006, \aap, 451, 99,
  \dodoi{10.1051/0004-6361:20054718}

\bibitem[{{Ben{\'\i}tez-Llambay} {et~al.}(2016){Ben{\'\i}tez-Llambay},
  {Navarro}, {Abadi}, {Gottl{\"o}ber}, {Yepes}, {Hoffman}, \&
  {Steinmetz}}]{2016MNRAS.456.1185B}
{Ben{\'\i}tez-Llambay}, A., {Navarro}, J.~F., {Abadi}, M.~G., {et~al.} 2016,
  \mnras, 456, 1185, \dodoi{10.1093/mnras/stv2722}

\bibitem[{{Cannon} {et~al.}(2012){Cannon}, {O'Leary}, {Weisz}, {Skillman},
  {Dolphin}, {Bigiel}, {Cole}, {de Blok}, \& {Walter}}]{2012ApJ...747..122C}
{Cannon}, J.~M., {O'Leary}, E.~M., {Weisz}, D.~R., {et~al.} 2012, \apj, 747,
  122, \dodoi{10.1088/0004-637X/747/2/122}

\bibitem[{{de Blok} \& {Walter}(2000)}]{2000ApJ...537L..95D}
{de Blok}, W.~J.~G., \& {Walter}, F. 2000, \apjl, 537, L95,
  \dodoi{10.1086/312777}

\bibitem[{{de Blok} \& {Walter}(2006)}]{2006AJ....131..343D}
---. 2006, \aj, 131, 343, \dodoi{10.1086/497829}

\bibitem[{{de Vaucouleurs} {et~al.}(1976){de Vaucouleurs}, {de Vaucouleurs}, \&
  {Corwin}}]{1976rcbg.book.....D}
{de Vaucouleurs}, G., {de Vaucouleurs}, A., \& {Corwin}, H.~G. 1976, {2nd
  reference catalogue of bright galaxies containing information on 4364
  galaxies with reference to papers published between 1964 and 1975}
  (University of Texas Press)

\bibitem[{{El-Badry} {et~al.}(2016){El-Badry}, {Wetzel}, {Geha}, {Hopkins},
  {Kere{\v{s}}}, {Chan}, \& {Faucher-Gigu{\`e}re}}]{2016ApJ...820..131E}
{El-Badry}, K., {Wetzel}, A., {Geha}, M., {et~al.} 2016, \apj, 820, 131,
  \dodoi{10.3847/0004-637X/820/2/131}

\bibitem[{{El-Badry} {et~al.}(2018){El-Badry}, {Quataert}, {Wetzel}, {Hopkins},
  {Weisz}, {Chan}, {Fitts}, {Boylan-Kolchin}, {Kere{\v{s}}},
  {Faucher-Gigu{\`e}re}, \& {Garrison-Kimmel}}]{2018MNRAS.473.1930E}
{El-Badry}, K., {Quataert}, E., {Wetzel}, A., {et~al.} 2018, \mnras, 473, 1930,
  \dodoi{10.1093/mnras/stx2482}

\bibitem[{{Faber} {et~al.}(2003){Faber}, {Phillips}, {Kibrick}, {Alcott},
  {Allen}, {Burrous}, {Cantrall}, {Clarke}, {Coil}, {Cowley}, {Davis}, {Deich},
  {Dietsch}, {Gilmore}, {Harper}, {Hilyard}, {Lewis}, {McVeigh}, {Newman},
  {Osborne}, {Schiavon}, {Stover}, {Tucker}, {Wallace}, {Wei}, {Wirth}, \&
  {Wright}}]{2003SPIE.4841.1657F}
{Faber}, S.~M., {Phillips}, A.~C., {Kibrick}, R.~I., {et~al.} 2003, in Society
  of Photo-Optical Instrumentation Engineers (SPIE) Conference Series, Vol.
  4841, \procspie, ed. M.~{Iye} \& A.~F.~M. {Moorwood}, 1657--1669,
  \dodoi{10.1117/12.460346}

\bibitem[{{Feast} {et~al.}(2012){Feast}, {Whitelock}, {Menzies}, \&
  {Matsunaga}}]{2012MNRAS.421.2998F}
{Feast}, M.~W., {Whitelock}, P.~A., {Menzies}, J.~W., \& {Matsunaga}, N. 2012,
  \mnras, 421, 2998, \dodoi{10.1111/j.1365-2966.2012.20525.x}

\bibitem[{{Friedli} {et~al.}(1994){Friedli}, {Benz}, \&
  {Kennicutt}}]{1994ApJ...430L.105F}
{Friedli}, D., {Benz}, W., \& {Kennicutt}, R. 1994, \apjl, 430, L105,
  \dodoi{10.1086/187449}

\bibitem[{{Gallart} {et~al.}(1996){Gallart}, {Aparicio}, {Bertelli}, \&
  {Chiosi}}]{1996AJ....112.2596G}
{Gallart}, C., {Aparicio}, A., {Bertelli}, G., \& {Chiosi}, C. 1996, \aj, 112,
  2596, \dodoi{10.1086/118205}

\bibitem[{{Ho} {et~al.}(2012){Ho}, {Geha}, {Munoz}, {Guhathakurta}, {Kalirai},
  {Gilbert}, {Tollerud}, {Bullock}, {Beaton}, \&
  {Majewski}}]{2012ApJ...758..124H}
{Ho}, N., {Geha}, M., {Munoz}, R.~R., {et~al.} 2012, \apj, 758, 124,
  \dodoi{10.1088/0004-637X/758/2/124}

\bibitem[{{Hopkins} {et~al.}(2014){Hopkins}, {Kere{\v{s}}}, {O{\~n}orbe},
  {Faucher-Gigu{\`e}re}, {Quataert}, {Murray}, \&
  {Bullock}}]{2014MNRAS.445..581H}
{Hopkins}, P.~F., {Kere{\v{s}}}, D., {O{\~n}orbe}, J., {et~al.} 2014, \mnras,
  445, 581, \dodoi{10.1093/mnras/stu1738}

\bibitem[{{Hubble}(1926)}]{1926ApJ....64..321H}
{Hubble}, E.~P. 1926, \apj, 64, 321, \dodoi{10.1086/143018}

\bibitem[{{Iorio} {et~al.}(2017){Iorio}, {Fraternali}, {Nipoti}, {Di Teodoro},
  {Read}, \& {Battaglia}}]{2017MNRAS.466.4159I}
{Iorio}, G., {Fraternali}, F., {Nipoti}, C., {et~al.} 2017, \mnras, 466, 4159,
  \dodoi{10.1093/mnras/stw3285}

\bibitem[{{Kacharov} {et~al.}(2017){Kacharov}, {Battaglia}, {Rejkuba}, {Cole},
  {Carrera}, {Fraternali}, {Wilkinson}, {Gallart}, {Irwin}, \&
  {Tolstoy}}]{2017MNRAS.466.2006K}
{Kacharov}, N., {Battaglia}, G., {Rejkuba}, M., {et~al.} 2017, \mnras, 466,
  2006, \dodoi{10.1093/mnras/stw3188}

\bibitem[{{Kaufmann} {et~al.}(2007){Kaufmann}, {Wheeler}, \&
  {Bullock}}]{2007MNRAS.382.1187K}
{Kaufmann}, T., {Wheeler}, C., \& {Bullock}, J.~S. 2007, \mnras, 382, 1187,
  \dodoi{10.1111/j.1365-2966.2007.12436.x}

\bibitem[{{Kawata} {et~al.}(2006){Kawata}, {Arimoto}, {Cen}, \&
  {Gibson}}]{2006ApJ...641..785K}
{Kawata}, D., {Arimoto}, N., {Cen}, R., \& {Gibson}, B.~K. 2006, \apj, 641,
  785, \dodoi{10.1086/500633}

\bibitem[{{Kirby} {et~al.}(2014){Kirby}, {Bullock}, {Boylan-Kolchin},
  {Kaplinghat}, \& {Cohen}}]{2014MNRAS.439.1015K}
{Kirby}, E.~N., {Bullock}, J.~S., {Boylan-Kolchin}, M., {Kaplinghat}, M., \&
  {Cohen}, J.~G. 2014, \mnras, 439, 1015, \dodoi{10.1093/mnras/stu025}

\bibitem[{{Kirby} {et~al.}(2013){Kirby}, {Cohen}, {Guhathakurta}, {Cheng},
  {Bullock}, \& {Gallazzi}}]{2013ApJ...779..102K}
{Kirby}, E.~N., {Cohen}, J.~G., {Guhathakurta}, P., {et~al.} 2013, \apj, 779,
  102, \dodoi{10.1088/0004-637X/779/2/102}

\bibitem[{{Komiyama} {et~al.}(2003){Komiyama}, {Okamura}, {Yagi}, {Furusawa},
  {Doi}, {Hamabe}, {Imi}, {Kimura}, {Miyazaki}, {Nakata}, {Okada}, {Ouchi},
  {Sekiguchi}, {Shimasaku}, {Yasuda}, {Arimoto}, \&
  {Ikuta}}]{2003ApJ...590L..17K}
{Komiyama}, Y., {Okamura}, S., {Yagi}, M., {et~al.} 2003, \apjl, 590, L17,
  \dodoi{10.1086/376551}

\bibitem[{{Leaman} {et~al.}(2017){Leaman}, {Mendel}, {Wisnioski}, {Brooks},
  {Beasley}, {Starkenburg}, {Martig}, {Battaglia}, {Christensen}, {Cole}, {de
  Boer}, \& {Wills}}]{2017MNRAS.472.1879L}
{Leaman}, R., {Mendel}, J.~T., {Wisnioski}, E., {et~al.} 2017, \mnras, 472,
  1879, \dodoi{10.1093/mnras/stx2014}

\bibitem[{{Letarte} {et~al.}(2002){Letarte}, {Demers}, {Battinelli}, \&
  {Kunkel}}]{2002AJ....123..832L}
{Letarte}, B., {Demers}, S., {Battinelli}, P., \& {Kunkel}, W.~E. 2002, \aj,
  123, 832, \dodoi{10.1086/338319}

\bibitem[{{Loebman} {et~al.}(2016){Loebman}, {Debattista}, {Nidever}, {Hayden},
  {Holtzman}, {Clarke}, {Ro{\v{s}}kar}, \& {Valluri}}]{2016ApJ...818L...6L}
{Loebman}, S.~R., {Debattista}, V.~P., {Nidever}, D.~L., {et~al.} 2016, \apjl,
  818, L6, \dodoi{10.3847/2041-8205/818/1/L6}

\bibitem[{{Mashchenko} {et~al.}(2008){Mashchenko}, {Wadsley}, \&
  {Couchman}}]{2008Sci...319..174M}
{Mashchenko}, S., {Wadsley}, J., \& {Couchman}, H.~M.~P. 2008, Science, 319,
  174, \dodoi{10.1126/science.1148666}

\bibitem[{{Massey} {et~al.}(2007){Massey}, {Olsen}, {Hodge}, {Jacoby},
  {McNeill}, {Smith}, \& {Strong}}]{2007AJ....133.2393M}
{Massey}, P., {Olsen}, K.~A.~G., {Hodge}, P.~W., {et~al.} 2007, \aj, 133, 2393,
  \dodoi{10.1086/513319}

\bibitem[{{Mayer} {et~al.}(2001){Mayer}, {Governato}, {Colpi}, {Moore},
  {Quinn}, {Wadsley}, {Stadel}, \& {Lake}}]{2001ApJ...559..754M}
{Mayer}, L., {Governato}, F., {Colpi}, M., {et~al.} 2001, \apj, 559, 754,
  \dodoi{10.1086/322356}

\bibitem[{{McConnachie}(2012)}]{2012AJ....144....4M}
{McConnachie}, A.~W. 2012, \aj, 144, 4, \dodoi{10.1088/0004-6256/144/1/4}

\bibitem[{{Navarro} {et~al.}(1996){Navarro}, {Frenk}, \&
  {White}}]{1996ApJ...462..563N}
{Navarro}, J.~F., {Frenk}, C.~S., \& {White}, S. D.~M. 1996, \apj, 462, 563,
  \dodoi{10.1086/177173}

\bibitem[{{Navarro} {et~al.}(1997){Navarro}, {Frenk}, \&
  {White}}]{1997ApJ...490..493N}
---. 1997, \apj, 490, 493, \dodoi{10.1086/304888}

\bibitem[{{P{\'e}rez} {et~al.}(2009){P{\'e}rez}, {S{\'a}nchez-Bl{\'a}zquez}, \&
  {Zurita}}]{2009A&A...495..775P}
{P{\'e}rez}, I., {S{\'a}nchez-Bl{\'a}zquez}, P., \& {Zurita}, A. 2009, \aap,
  495, 775, \dodoi{10.1051/0004-6361:200811084}

\bibitem[{{Postman} \& {Geller}(1984)}]{1984ApJ...281...95P}
{Postman}, M., \& {Geller}, M.~J. 1984, \apj, 281, 95, \dodoi{10.1086/162078}

\bibitem[{{Sandage}(2005)}]{2005ARA&A..43..581S}
{Sandage}, A. 2005, \araa, 43, 581,
  \dodoi{10.1146/annurev.astro.43.112904.104839}

\bibitem[{{Spekkens} {et~al.}(2014){Spekkens}, {Urbancic}, {Mason}, {Willman},
  \& {Aguirre}}]{2014ApJ...795L...5S}
{Spekkens}, K., {Urbancic}, N., {Mason}, B.~S., {Willman}, B., \& {Aguirre},
  J.~E. 2014, \apjl, 795, L5, \dodoi{10.1088/2041-8205/795/1/L5}

\bibitem[{{Swan} {et~al.}(2016){Swan}, {Cole}, {Tolstoy}, \&
  {Irwin}}]{2016MNRAS.456.4315S}
{Swan}, J., {Cole}, A.~A., {Tolstoy}, E., \& {Irwin}, M.~J. 2016, \mnras, 456,
  4315, \dodoi{10.1093/mnras/stv2774}

\bibitem[{{Thompson} {et~al.}(2016){Thompson}, {Ryan}, \&
  {Sibbons}}]{2016MNRAS.462.3376T}
{Thompson}, G.~P., {Ryan}, S.~G., \& {Sibbons}, L.~F. 2016, \mnras, 462, 3376,
  \dodoi{10.1093/mnras/stw1193}

\bibitem[{{Valenzuela} {et~al.}(2007){Valenzuela}, {Rhee}, {Klypin},
  {Governato}, {Stinson}, {Quinn}, \& {Wadsley}}]{2007ApJ...657..773V}
{Valenzuela}, O., {Rhee}, G., {Klypin}, A., {et~al.} 2007, \apj, 657, 773,
  \dodoi{10.1086/508674}

\bibitem[{{Walker} \& {Pe{\~n}arrubia}(2011)}]{2011ApJ...742...20W}
{Walker}, M.~G., \& {Pe{\~n}arrubia}, J. 2011, \apj, 742, 20,
  \dodoi{10.1088/0004-637X/742/1/20}

\bibitem[{{Weisz} {et~al.}(2014){Weisz}, {Dolphin}, {Skillman}, {Holtzman},
  {Gilbert}, {Dalcanton}, \& {Williams}}]{2014ApJ...789..147W}
{Weisz}, D.~R., {Dolphin}, A.~E., {Skillman}, E.~D., {et~al.} 2014, \apj, 789,
  147, \dodoi{10.1088/0004-637X/789/2/147}

\bibitem[{{Weldrake} {et~al.}(2003){Weldrake}, {de Blok}, \&
  {Walter}}]{2003MNRAS.340...12W}
{Weldrake}, D.~T.~F., {de Blok}, W.~J.~G., \& {Walter}, F. 2003, \mnras, 340,
  12, \dodoi{10.1046/j.1365-8711.2003.06170.x}

\bibitem[{{Wheeler} {et~al.}(2017){Wheeler}, {Pace}, {Bullock},
  {Boylan-Kolchin}, {O{\~n}orbe}, {Elbert}, {Fitts}, {Hopkins}, \&
  {Kere{\v{s}}}}]{2017MNRAS.465.2420W}
{Wheeler}, C., {Pace}, A.~B., {Bullock}, J.~S., {et~al.} 2017, \mnras, 465,
  2420, \dodoi{10.1093/mnras/stw2583}

\bibitem[{{Wolf} {et~al.}(2010){Wolf}, {Martinez}, {Bullock}, {Kaplinghat},
  {Geha}, {Mu{\~n}oz}, {Simon}, \& {Avedo}}]{2010MNRAS.406.1220W}
{Wolf}, J., {Martinez}, G.~D., {Bullock}, J.~S., {et~al.} 2010, \mnras, 406,
  1220, \dodoi{10.1111/j.1365-2966.2010.16753.x}

\bibitem[{{Wyder}(2001)}]{2001AJ....122.2490W}
{Wyder}, T.~K. 2001, \aj, 122, 2490, \dodoi{10.1086/323706}

\bibitem[{{Zhuang} {et~al.}(2019){Zhuang}, {Leaman}, {van de Ven}, {Zibetti},
  {Gallazzi}, {Zhu}, {Falc{\'o}n-Barroso}, \&
  {Lyubenova}}]{2019MNRAS.483.1862Z}
{Zhuang}, Y., {Leaman}, R., {van de Ven}, G., {et~al.} 2019, \mnras, 483, 1862,
  \dodoi{10.1093/mnras/sty2916}

\end{thebibliography}
\bibliographystyle{aasjournal}

\end{document}